\documentclass[a4paper,12pt]{article}

\date{}

\usepackage{bm}
\usepackage[dvips]{graphicx}
\title{Spectator isobar production in\\ the $A\left( {\gamma ,\pi NN} \right)B$ reaction}
\author{ I.V. Glavanakov\footnote{E-mail: glavanakov@tpu.ru}, 
A.N. Tabachenko\footnote{E-mail: ant1936@tpu.ru}\\
\it Institute of Physics and Technology, Tomsk Polytechnic University,\\
\it Tomsk, Russia}
\begin{document}
\maketitle

\bigskip
\begin{abstract}
We present an analysis of the spectator mechanism of $\Delta $-isobar 
production in the pion photoproduction on nuclei with the emission of two 
nucleons. The reaction mechanism is studied within the framework of the 
$\Delta N$-correlation model, which considers the isobar and nucleon of the 
$\Delta N$-system produced in the nucleus at the virtual $NN \to \Delta N$ 
transition, to be in a dynamic relationship. The two-particle transition 
operator for nuclei is obtained by the \textit{S}-matrix approach. We 
consider the properties of the spectator mechanism of isobar production 
using the example of the reaction $^{16}$O$\left( {\gamma ,\ \pi ^{ - 
}pn} \right){}^{14}$O. Numerical estimates of the cross section are obtained 
in the kinematic region, where it is possible to expect the manifestation of 
bound isobar-nuclear states.\\[5mm]
\end{abstract}

\section{Introduction}
In the impulse approximation the amplitude of the nuclear reaction $A\left( 
{a,a'} \right)bC$, accompanied by the two-particle fragmentation of the 
nucleus \textit{A}, is the sum of two terms. Each term describes the 
reaction mechanism, in which the incident particle interacts with one of the 
fragments and the other nuclear fragment (spectator) does not participate in 
that interaction. The terms describing the interaction of a particle with 
light and heavy fragments are usually called the direct and exchange 
amplitudes, respectively \cite{Goldberger}. Data from nuclear reactions of the type 
$A\left( {e,e'p} \right)B$ and $A\left( {p,2p} \right)B$, measured in the 
kinematic region where the direct amplitude dominates, are an important 
source of information on single-particle properties of the nuclear shell 
structure \cite{Jacob,Frullani}. The kinematic region of the exchange reaction mechanism is, 
as a rule, outside the measurement capabilities of experimental setups, in 
which the identification of the reaction events is carried out by detecting 
two fast particles. An exception to this rule is the case when the spectator 
is a nucleon resonance (isobar).

The isobar as spectator is possible in the framework of a model in which the 
wave function of the nucleus includes both nucleon and isobar 
configurations. According to this model, some nucleons in the ground state 
of the nucleus can experience internal excitation and go through a process 
$NN \to \Delta N$ resulting in a virtual isobar state \cite{Green,Weber,Frick}.
Such isobars 
can be knocked out of the nucleus by a high-energy particle. The isobar 
knock-out is possible both in the direct interaction of the incident 
particle with the isobar, and as a result of the exchange (spectator) 
reaction mechanism, in which the incident particle interacts with the 
nucleon core of the nucleus. The isobar, which did not participate in the 
interaction, goes into a free state as a result of "shaking". 

Direct isobar knock-out from nucleus has been studied both experimentally 
and theoretically, see for example \cite{Amelin,Jonsson,Gerasimov,Glavanakov:2013}. 
The spectator mechanism of isobar 
production has been analysed until recently only in reactions on the 
lightest nuclei \cite{Jonsson,Kisslinger,Nath}.

The purpose of studying isobar configurations in the ground state of nuclei 
is to obtain information, on the one hand, about the high-momentum component 
of the wave function of the atomic nucleus, and on the other hand, about the 
behavior of $\Delta $-isobars in a nuclear medium. One interesting aspect of 
$\Delta $-nuclear physics is the question of the existence of bound 
isobar-nuclear states, the states of the nucleus in which one of the 
nucleons is replaced by a ``real'' isobar (near the mass shell). The 
excitation energy of such nuclei ($\Delta $-nuclei) can be $\sim $ 300 MeV. 
Decay of a $\Delta $-nuclei is possible by emission of a pion-nucleon pair, 
or two nucleons as a result of the transition $\Delta N \to NN$.

One of the manifestations of a $\Delta $-nuclei in the nuclear reaction is 
the detection of a pion and a nucleon, which fly out in opposite directions 
with a small total momentum. Based on such features, and assuming the 
existence of a $\Delta $-nuclei, the data of the reactions measured in 
experiments on the MAMI microtron in Mainz \cite{Bartsch} and the Tomsk 
synchrotron \cite{Glavanakov:2008} were interpreted. Some signs of the $\Delta $-nuclei 
are also observed in data from other experiments \cite{Glavanakov:2009}. It should be 
noted that theoretical estimates of the possibility of the existence of a 
$\Delta $-nuclei are contradictory \cite{Arenhovel,Belyaev,Mota}.

One of the problems associated with identifying such exotic nuclear states 
is the determination of the reaction mechanisms that could imitate them. 
From general physical concepts, it follows that the spectator $\Delta 
$-isobar production can imitate the excitation of a $\Delta $-nucleus. This 
mechanism of isobar production was proposed in the Ref. \cite{Glavanakov:2017}. 
The aim of 
this paper is to develop further this model for the isobar production and to 
study its properties in the kinematic region, where it is possible to expect 
the manifestation of bound isobar-nuclear states. 

The paper is arranged as follows: In Sec. 2 the transition matrix for the 
spectator mechanism of $\Delta $-isobar production in the $A\left( {\gamma 
,\pi NN} \right)B$ reaction is given. The two-particle transition operator 
is obtained in Sec. 3. Compared with Ref. \cite{Glavanakov:2017} we start from the 
relativistic approach for the working-out of the transition operator. The 
detail analysis of the spectator mechanism of $\Delta $-isobar production in 
the ${}^{16}$O$\left( {\gamma ,\quad \pi ^{ -} pn} \right){}^{14}$O reaction 
is made in Sec. 4. The analysis of the reaction mechanism is performed 
within the framework of the $\Delta N$-correlation model \cite{Glavanakov:2013}.

\section{Nuclear transition matrix}

The differential cross section for the $A\left( {\gamma ,^{}\pi NN} 
\right)B$ reaction is written in the laboratory system of coordinates as 
\begin{eqnarray*}
d\sigma & = & \left( {2\pi}  \right)^{4}\delta \left( {E_{\gamma}  + M_{T} - 
E_{\pi}  - E_{n1} - E_{n2} - E_{B}}  \right)\delta \left({{\bf p}_{\gamma}  - 
{\bf p}_{\pi}  - {\bf p}_{n1} - {\bf p}_{n2} - {\bf p}_{B}}  \right)\\
& \times &  \left|{T_{fi}}\right|^{2}
\frac{{d^{3}p_{\pi} } }{{\left( {2\pi}  
\right)^{3}}}\frac{{d^{3}p_{n1}} }{{\left( {2\pi}  
\right)^{3}}}\frac{{d^{3}p_{n2}} }{{\left( {2\pi}  
\right)^{3}}}\frac{{d^{3}p_{B}} }{{\left( {2\pi}  \right)^{3}}},
\end{eqnarray*}
where $\left( {E_{\gamma}  ,{\bf p}_{\gamma} }  \right),\ \left( {E_{\pi}  
,{\bf p}_{\pi} }  \right),\ \left( {E_{n1} ,{\bf p}_{n1}}  \right),\ \left( 
{E_{n2} ,{\bf p}_{n2}}  \right),\ \left( {E_{B} ,{\bf p}_{B}}  \right)$ are 
4-momenta of the photon, pion, two free nucleons, and the residual nucleus 
\textit{B}, $M_{T} $ is the mass of the nucleus \textit{A}, and $T_{fi} $ is 
the transition matrix.

Within the framework of the formalism developed in Ref. \cite{Glavanakov:2013}, the matrix of the 
transition from the initial state containing the photon and the nucleus 
\textit{A}, to the final state including the pion, two free nucleons, and 
the residual nucleus \textit{B,} is represented as follows:
\begin{eqnarray}\label{eq1}
 T_{fi}& = &\frac{{A\left( {A - 1} \right)}}{{2}}\int {d\left( {X_{1} ',X_{2} 
'X_{1} ,X_{2} ,...,X_{A}}  \right)_{}} \nonumber \\
  &\times & \Psi _{F}^{\ast}  \left( {X_{1} ',X_{2} ',X_{3} ,...,X_{A}}  \right) < 
X_{1} ',X_{2} '|t_{\gamma \pi}  |X_{1} ,X_{2} > \Psi _{i} \left( {X_{1} 
,X_{2} ,X_{3} ,...,X_{A}}  \right). 
\end{eqnarray}
Here, $\Psi _{i} $ is the wave function of the nucleus \textit{A}, in which 
some of the baryons are in the isobar state, $\Psi _{F} $ is the wave 
functions of the finite nuclear system, which includes the two free nucleons 
and the residual nucleus \textit{B,} $X \equiv \left( {x,m} \right) \equiv 
\left( {{\bf r},s,t,m} \right)$ is the complete set of coordinates characterising 
the position of the baryon in the ordinary space (\textbf{r}), spin space 
(\textit{s}), isotopic space (\textit{t}), and in the space of internal 
states (\textit{m}), $t_{\gamma \pi}  $ is a two-body operator for pion 
production from nuclei with the emission of two nucleons, and the integral 
sign denotes integration over space variables and summation over spin and 
isotopic variables and over the variables of the internal state of baryons. 
In the framework of this approach, the wave function of the system of 
\textit{A} baryons is written in the form \cite{Horlacher}
$$
\Psi \left( {X_{1} ,...,X_{A}}  \right) = A_{m} \ \phi _{N_{1} 
...N_{A}}  \left( {m_{1} ,...,m_{A}}  \right)\ \psi ^{N_{1} ...N_{A} 
}\left( {x_{1} ,...,x_{A}}  \right),
$$
where $\phi _{N_{1} N_{2} ...N_{A}}  \left( {m_{1} ,m_{2} ,...,m_{A}}  
\right)$ is the function describing the internal states of baryons. The 
state indices take the values \textit{N} (nucleon) or $\Delta $ (isobar). 
The wave function describing the state of baryons in the space (\textbf{r}) 
and spin (\textit{s}) and isotopic (\textit{t}) spaces is $\psi ^{N_{1} 
N_{2} ...N_{A}} \left( {x_{1} ,x_{2} ,...,x_{A}}  \right)$, and $A_{m} $ is 
the antisymmetrization operator. The operators of the pion production 
$t_{\gamma \pi}  $, acting in the spaces of the coordinates \textit{x} and 
\textit{X}, are related by the equation
$$
< x_{1} ',x_{2} '|t_{\gamma \pi}  |x_{1} ,x_{2} > ^{} = 
^{}\sum\limits_{m_{1} ',m_{2} ',m_{1} ,m_{2}}  {\phi _{NN}^{\ast}  \left( 
{m_{1} ',m_{2} '} \right) < X_{1} ',X_{2} '|t_{\gamma \pi}  |X_{1} ,X_{2} > 
\phi _{N\Delta}  \left( {m_{1} ,m_{2}}  \right)}.
$$

	According to the $\Delta N$-correlation model, the wave function $\Psi _{i} 
$ of the initial nucleus \textit{A} is represented as an antisymmetrized 
product of the wave function $\Psi _{\left[ {\beta _{i} \beta _{j}}  
\right]}^{N\Delta}  $ of the $\Delta N$ system, which includes the isobar 
and nucleon produced at the virtual transition $NN \to \Delta N$ of two 
nucleons in states $\beta _{i} $ and $\beta _{j} $, and the wave function 
$\Psi _{\left( {\beta _{i} \beta _{j}}  \right)^{ - 1}} $ of the nucleon 
core \cite{Horlacher}:
\begin{equation}\label{eq2}
\Psi _{i} \left( {X_{1} ,...,X_{A}}  \right)^{} = ^{}A_{12;3...A}^{} 
\sum\limits_{ij} {\Psi _{\left[ {\beta _{i} \beta _{j}}  \right]}^{N\Delta}  
\left( {X_{1} ,X_{2}}  \right)^{}\Psi _{\left( {\beta _{i} \beta _{j}}  
\right)^{ - 1}}^{} \left( {X_{3} ,...,X_{A}}  \right)} , 	
\end{equation}		
where $A_{12;3...A}^{} $ is an antisymmetrization operator.

The matrix element $T_{fi} $ (\ref{eq1}) with the wave function  
$\Psi _{i}^{} $ (\ref{eq2}) is the sum of terms, each of which 
corresponds to a certain reaction mechanism. Part of the possible mechanisms 
of the $A\left( {\gamma ,^{}\pi NN} \right)B$ reaction, for which the 
initiating processes were $\gamma \Delta \to N\pi $ and $\gamma N \to N\pi 
$, were analysed in Ref. \cite{Glavanakov:2013}. 
We will now consider the mechanisms of the 
reaction in which a photon is absorbed by a nucleus, with the initiating 
process being a transition $\gamma N \to N$.

We will assume that the operator $t_{\gamma \pi}  $ describes the absorption 
of a photon by a nucleon 1, and that nucleon 2 appears as a result of the 
decay of the isobar $\Delta \to N\pi $. Then, the spectator mechanism of the 
isobar production corresponds to the following components of $T_{fi} $ 
$$
	T_{sp} = T_{1} + T_{2} + T_{3} .
$$
Here,
$$
\displaylines{
 T_{1} = - \left( {A - 2} \right)\int {d\left( {X_{1} ',X_{2} 'X_{1} ,X_{2} 
,...,X_{A}}  \right)_{}}\  \varphi _{\alpha _{1} \alpha _{2}} ^{\ast}  \left( 
{X_{1} ',X_{2} '} \right)^{}\ \times \cr 
 \Psi _{f}^{\ast}  \left( {X_{3} ,X_{4} ,...,X_{A}}  \right) < X_{1} ',X_{2} 
'|t_{\gamma \pi}  |X_{1} ,X_{2} > \sum\limits_{ij} {} \Psi _{\left[ {\beta 
_{i} \beta _{j}}  \right]}^{N\Delta}  \left( {X_{3} ,X_{2}}  \right)^{}\Psi 
_{\left( {\beta _{i} \beta _{j}}  \right)^{ - 1}}^{} \left( {X_{1} ,X_{4} 
,...,X_{A}}  \right),}
$$
\vspace*{-0.8cm}
\[
\displaylines{
 T_{2} = \int {d\left( {X_{1} ',X_{2} 'X_{1} ,X_{2} ,...,X_{A}}  \right)_{} 
}\ \varphi _{\alpha _{1} \alpha _{2}} ^{\ast}  \left( {X_{1} ',X_{2} '} 
\right)^{} \ \times \cr 
 \Psi _{f'}^{\ast}  \left( {X_{3} ,X_{4} ,...,X_{A}}  \right) < X_{1} 
',X_{2} '|t_{\gamma \pi}  |X_{1} ,X_{2} > \quad \sum\limits_{ij} {} \Psi 
_{\left[ {\beta _{i} \beta _{j}}  \right]}^{N\Delta}  \left( {X_{1} ,X_{2}}  
\right)^{}\Psi _{\left( {\beta _{i} \beta _{j}}  \right)^{ - 1}}^{} \left( 
{X_{3} ,X_{4} ,...,X_{A}}  \right), \cr} 
\]
\vspace*{-0.8cm}
\[
\displaylines{
 T_{3} = - \left[ {\frac{{A\left( {A - 1} \right)}}{{2}}} \right]^{{{1} 
\mathord{\left/ {\vphantom {{1} {2}}} \right. \kern-\nulldelimiterspace} 
{2}}}\left( {A - 2} \right)\int {d\left( {X_{1} ',X_{2} 'X_{1} ,X_{2} 
,...,X_{A}}  \right)_{}} \ \varphi _{\alpha _{n1} \alpha _{n2}} ^{\ast}  \left( 
{X_{3} ,X_{2} '} \right)^{} \ \times \cr 
 \Psi _{f"}^{\ast}  \left( {X_{1} ',X_{4} ,...,X_{A}}  \right) < X_{1} 
',X_{2} '|t_{\gamma \pi}  |X_{1} ,X_{2} > \Psi _{i} \left( {X_{1} ,X_{2} 
,X_{3} ,...,X_{A}}  \right), \cr} 
\]
\noindent
where $\varphi _{\alpha _{1} \alpha _{2}} ^{} $ is the antisymmetric wave 
function of the two free nucleons in states $\alpha _{n1} $ and $\alpha 
_{n2} $, $\Psi _{f\left( {',"} \right)}^{} $ is the wave function describing 
the state of the residual nucleus \textit{B}. 

\vspace{-1.5cm}
\begin{figure}[ph]
\unitlength=1cm
\centering
\begin{picture}(4,4)
\put(-1,0){\includegraphics[width=4cm,keepaspectratio]{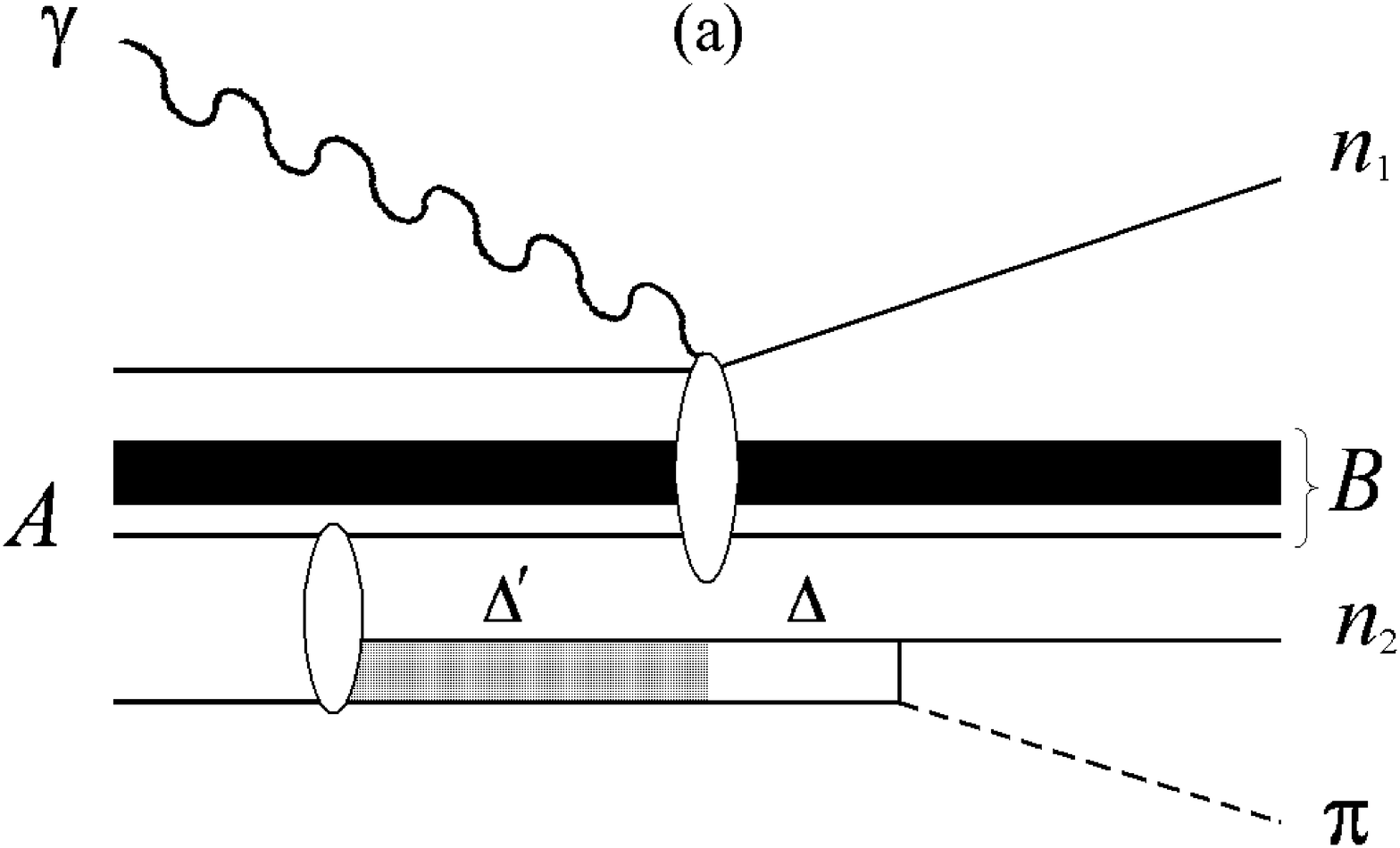}}
\end{picture}
\begin{picture}(4,4)
\put(0,0.5){\includegraphics[width=4cm,keepaspectratio]{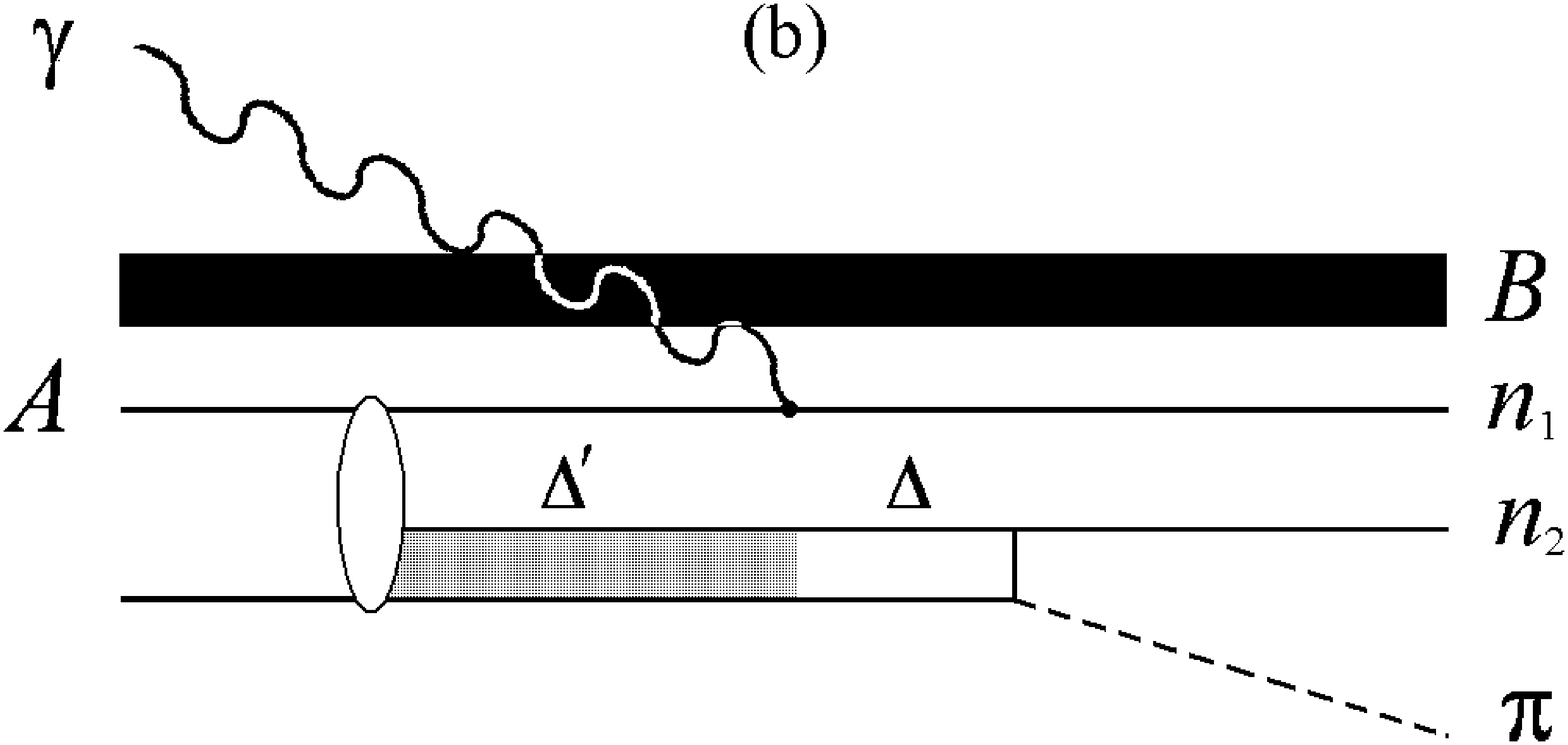}}
\end{picture}
\begin{picture}(4,4)
\put(1,0){\includegraphics[width=4cm,keepaspectratio]{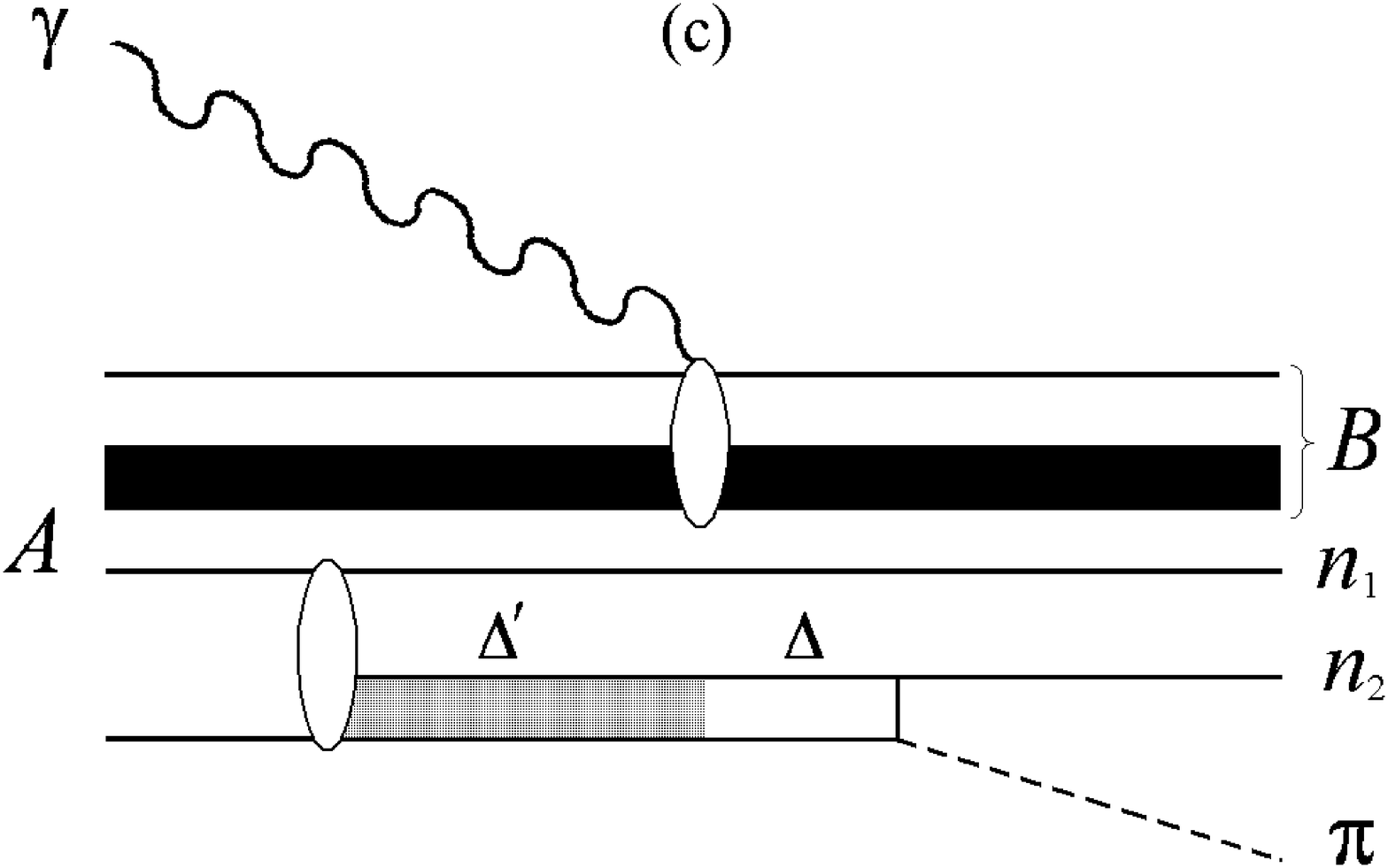}}
\end{picture}
\caption{\small Diagrams illustrating the spectator mechanisms of $\Delta $-isobar 
production in the $A\left( {\gamma ,_{} \pi NN} \right)B$ reaction.
\protect\label{fig1}}
\end{figure}
The diagrams in Figs. \ref{fig1}\textit{a}, \ref{fig1}\textit{b}, and \ref{fig1}\textit{c} 
illustrate the reaction mechanisms corresponding to the amplitudes\textit{ T}$_{1}$, 
\textit{T}$_{2,}$ and \textit{T}$_{3.}$ The reaction mechanisms corresponding 
to these amplitudes differ by the state of the nucleon 
that absorbed the photon. In \textit{T}$_{1}$ and \textit{T}$_{2}$ nucleon 1 
goes into a free state, and in \textit{T}$_{3}$ it remains in the nucleus in 
the bound state. Regardless of the reaction mechanism for knock-out of a 
virtual isobar from the nucleus, it is necessary that the incident particle 
transmit energy to the nucleus of about 300 MeV. This is accompanied by a 
transfer to the nucleus of momentum, the heavier the incident particle, the 
greater the magnitude of the momentum transfer. Even in the case of the 
$A\left( {\gamma ,^{}\pi NN} \right)B$ reaction, it is unlikely that the 
nucleon that absorbed the photon remains in the nucleus in the bound state. 
Therefore, the amplitude \textit{T}$_{3}$ can be neglected. We will not 
consider also amplitude \textit{T}$_{2}$ which value is $ \sim \left( {A - 
2} \right)$ times less than amplitude \textit{T}$_{1.}$  In the case of amplitude 
\textit{T}$_{1}$, on absorption of the photon by the nucleus, the "active" 
nucleon passes into a free state with a momentum $p_{n1} $, the nucleon core 
receives a momentum equal to $p_{\gamma}  - p_{n1} $. As a result, the 
virtual isobar becomes real and decays into a pion and nucleon.

\section{Transition operator}
 
The two-particle transition operator $t_{\gamma \pi}$ for nuclei was 
obtained following the \textit{S}-matrix approach developed in Ref. \cite{Chemtob}. 
The matrix element of the \textit{S}-matrix, in the lowest order of perturbation 
theory, of the elemental process, where the $\Delta $-isobar and nucleon, 
which absorbs a photon, pass into two nucleons with an emission of a pion, 
is written in the interaction representation as
\begin{equation}\label{eq3}
S_{fi} = \ < \alpha _{\pi}  ,\alpha _{n1} ,\alpha _{n2} |\int 
{d^{4}r \ }  A_{\mu}  \left( {r} \right)\ J^{\mu} \left( {r} 
\right)|\alpha _{\gamma}  ,\alpha _{n} ,\alpha _{\Delta}  > .		
\end{equation} 
Here, $r \equiv \left( {{\bf r},t} \right)$,\textit{ t} is time, $\alpha _{\gamma 
} ,\alpha _{\pi}  ,\alpha _{n} ,\alpha _{\Delta}  $ are the state indices of 
the photon, pion, initial nucleon, and isobar, and $A_{\mu}  \left( {r} 
\right)$ is the 4-potential of the electromagnetic field,
$$
J^{\mu} \left( {r} \right) = i\int {d^{4}r_{1}}  \ d^{4}r_{2} \ 
T\left( {j^{\mu} \left( {r} \right)L_{s}^{1} \left( {r_{1}}  
\right)L_{s}^{2} \left( {r_{2}}  \right)} \right),
$$
where $j^{\mu} \left( {r} \right)$ is electromagnetic current of the 
nucleons, the effective Lagrangian $L_{s}^{1} \left( {r_{1}}  
\right)$ describes the interaction of the nucleon, isobar, and pion, 
$L_{s}^{2} \left( {r_{2}}  \right)$ effectively describes the transition of 
the isobar from the virtual state to the real state, and \textit{T} is the 
time-ordering operator.

The electromagnetic current of the nucleons is written as
$$
j_{\mu}  \left( {r} \right) = \ - \ e\ \bar {\psi} _{N} \left( 
{r} \right)\ \gamma _{\mu}  \ I_{e} \ \psi _{N} \left( {r} 
\right) + \frac{{e}}{{2M_{N}} }\ \frac{{\partial} }{{\partial r_{\nu}  
}}\left( {\bar {\psi} _{N} \left( {r} \right)\ I_{m} \ \sigma _{\mu 
\nu}  \ \psi _{N} \left( {r} \right)} \right),
$$
where \textit{e} is the electron charge, $\psi _{N} $ is the nucleon field 
operator, $\gamma _{\mu}$ is Dirac matrix, $\sigma _{\mu \nu}  = \left( {\gamma _{\mu}  \gamma _{\nu}  - 
\gamma _{\nu}  \gamma _{\mu} }  \right)/2$, and
$$
I_{e} = \frac{{1 + \tau _{3}} }{{2}},\quad \quad I_{m} = \frac{{\mu _{p} + 
\mu _{n}} }{{2}} + \frac{{\mu _{p} - \mu _{n}} }{{2}}\tau _{3} .
$$
Here, $\mu _{p} ,\ \mu _{n} $ are the magnetic moments of the proton and 
neutron in Bohr magnetons, $\tau _3$ is Pauli matrix.

The effective Lagrangian $L_{s}^{1} \left( {r} \right)$, describes the 
transition $\Delta \to N\pi $
$$
L_{s}^{1} \left( {r} \right) = \ g_{\Delta N\pi}  \ \frac{{\partial 
}}{{\partial r_{\nu} } }\ \boldsymbol{\phi} _{\pi}  \left( {r} \right)\ \bar 
{\psi} _{N} \left( {r} \right)\ \Gamma _{\nu \mu} ^{\ast}  \ {\bf T}^{ + 
}\ \psi _{\Delta} ^{\mu}  \left( {r} \right),
$$
where $g_{\Delta N\pi}  $ is the $\pi N \Delta$ coupling constant, 
$\Gamma _{\mu \nu}  = \left( {g_{\mu \nu}  + C\ 
\gamma _{\mu}  \gamma _{\nu} }  \right), \ C = - 1/4$ \cite{Peccei}, ${\bf T}^{ +} $ is the 
transition operator between states with isospins 3/2 and 1/2, $\boldsymbol{\phi} _{\pi}  
$ is the pion field operator, and $\psi _{\Delta} ^{\mu}  $ is the field 
operator of a real isobar with spin and isospin 3/2. The effective 
Lagrangian describing the transition of the isobar from the virtual state to 
the real state is used in the form
$$
L_{s}^{2} \left( {r} \right) = g\ \left( {\bar {V}_{\Delta \mu}  \left( 
{r} \right)\ \psi _{\Delta} ^{\mu}  \left( {r} \right) + \bar {\psi 
}_{\Delta \mu}  \left( {r} \right)\ V_{\Delta} ^{\mu}  \left( {r} 
\right)} \right),
$$
where \textit{g} is the transition constant (equal to unity in the spectator 
model), and $V_{\Delta} ^{\mu}  $ is the operator of the virtual isobar 
field.

\begin{figure}[ph]
\centerline{\includegraphics[width=2.0in]{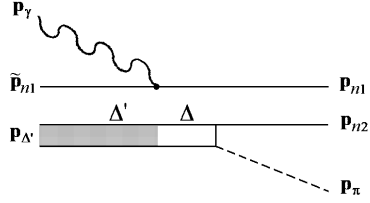}}
\vspace*{8pt}
\caption{\small Diagram representing the matrix element of the \textit{S} matrix 
corresponding to the spectator mechanism of isobar production.
\protect\label{fig2}}
\end{figure}

Reducing the \textit{T}-product of the operators in (\ref{eq3}) to the normal form, 
we obtain the sum of the terms whose matrix elements describe definite 
physical processes. The matrix element of the two-particle transition 
operator in momentum space, which corresponds to the spectator mechanism of 
the reaction, is determined by the matrix element of the \textit{S} matrix, 
represented by the Feynman diagram in Fig. \ref{fig2}. The representation of the 
matrix element of the transition operator in configuration space $S_{sp} $, 
obtained by doing a Fourier transform on each baryon momentum, is given by
$$
\displaylines{
 S_{sp} = \delta \left( {{\bf r}_{1}^{'} - {\bf r}_{1}}  \right)\ \delta \left( 
{{\bf r}_{2}^{'} - {\bf r}_{2}}  \right)\ \times \cr 
 \sqrt {\frac{{M_{N}} }{{E_{n1}} }\ }  \bar {u}_{m_{\sigma 1}}  \left( 
{{\bf p}_{n1}}  \right)\ \xi _{m_{\tau 1}} ^{ +}  \ \Gamma _{\mu} ^{\gamma 
} \left( {{\bf p}_{\gamma} }  \right)\ \sqrt {\frac{{M_{N}} }{{\tilde {E}_{n1} 
}}} \ u_{\tilde {m}_{\sigma 1}}  \left( {\tilde {{\bf p}}_{n1}}  \right)\ 
\xi _{\tilde {m}_{\tau 1}}  \frac{{e^{i{\bf p}_{\gamma}  {\bf r}_{1}} }}{{\sqrt 
{2E_{\gamma} } } }\ {\varepsilon}_{}^{\mu}  \left( {{\bf p}_{\gamma} }  
\right)\ \times \cr 
 \frac{{e^{ - i{\bf p}_{\pi}  {\bf r}_{2}} }}{{\sqrt {2E_{\pi} } } }\ \sqrt 
{\frac{{M_{N}} }{{E_{n2}} }} \ \bar {u}_{m_{\sigma 2}}  \left( {{\bf p}_{n2}}  
\right)\ \xi _{m_{\tau 2}} ^{ +}  \ \Gamma _{\nu} ^{\Delta}  \left( 
{{\bf p}_{\pi}  ,{\bf p}_{\Delta} }  \right)\ \sqrt {\frac{{M_{\Delta '} 
}}{{E_{\Delta '}} }\ }  u_{m_{\sigma \Delta} } ^{\nu}  \left( {{\bf p}_{\Delta 
}}  \right)\ \xi _{m_{\tau \Delta} }  . \cr} 
$$ 
Here, $M_{N} $ and $M_{\Delta '} $ are the masses of the nucleon and virtual 
isobar, $u_{m} \left( {\bf p} \right)$ and $u_{m}^{\nu}  \left( {\bf p} \right)$ are 
Dirac and Rarita-Schwinger spinors normalized as $\bar {u}u = 1$, $\xi 
_{m_{\tau 1\left( {2} \right)}} ^{ +}  $ and $\xi _{m_{\tau \Delta} }  $ are 
isotopic spinors of the nucleons and isobar, ${\varepsilon} _{}^{\mu}  \left( 
{{\bf p}_{\gamma} }  \right)$ is the photon polarization 4-vector, ${\bf p}_{\Delta}  = 
{\bf p}_{\pi}  + {\bf p}_{n2} ,$
$$
\Gamma _{\mu} ^{\gamma}  \left( {{\bf p}_{\gamma} }  \right) = - \ e\ 
\left( {\gamma _{\mu}  \ I_{e} + \frac{{i}}{{2M_{N}} }\ {p}_{\gamma} ^{\nu}  
\ \sigma _{\mu \nu}  \ I_{m}}  \right), \ \Gamma _{\nu} ^{\Delta}  
\left( {{\bf p}_{\pi}  ,{\bf p}_{\Delta} }  \right) =  - \ {p}_{\pi} ^{\tilde {\nu} } \ 
\Gamma _{\tilde {\nu} \mu} ^{\ast}  \ S_{\nu} ^{\mu}  \left( {{\bf p}_{\Delta 
}}  \right)\ \boldsymbol{\phi} _{\pi}  \cdot {\bf T}^{ +} ,
$$
and $S_{\nu} ^{\mu}  \left( {{\bf p}_{\Delta} }  \right)$ is the $\Delta $-isobar 
propagator \cite{Williams}.

Passing to the nonrelativistic limit, accurate to terms on the order of 
$O\left( {p/M} \right)$, the operator $t_{\gamma \pi}  $ was obtained in the 
form
$$
\displaylines{
 t_{\gamma \pi}  = \delta \left( {{\bf r}_{1}^{'} - {\bf r}_{1}}  \right)\ \delta 
\left( {{\bf r}_{2}^{'} - {\bf r}_{2}}  \right)\ \times \cr 
 \ \ \frac{{e^{ - i{\bf p}_{\pi}  {\bf r}_{2}} }}{{\sqrt {2E_{\pi} } } }\ 
t_{\Delta N}^{S} \ t_{\Delta N}^{I} \frac{{2M_{\Delta} } }{{\left( 
{E_{n2} + E_{\pi} }  \right)^{2} + \left( {{\bf p}_{n2} + {\bf p}_{\pi} }  \right)^{2} - 
M_{\Delta} ^{2} + i\ \Gamma _{\Delta}  M_{\Delta} } }\left( {t_{\gamma 
N}^{Se} \ t_{\gamma N}^{Ie} + t_{\gamma N}^{Sm} t_{\gamma N}^{Im}}  
\right)\ \frac{{e^{i{\bf p}_{\gamma}  {\bf r}_{1}} }}{{\sqrt {2E_{\gamma} } } }, 
} 
$$ 
where
$$
\displaylines{
 t_{\gamma N}^{Se} \left( {{\bf p}_{\gamma}  ,\bm{\varepsilon}}  \right) = 
\frac{{i\ e}}{{2M_{N}} }{\bm \sigma} \cdot {\bf p}_{\gamma}  \ {\bm \sigma} \cdot 
{\bm \varepsilon} ,\ \ \ t_{\gamma N}^{Ie} = ^{}\frac{{1 + \tau 
_{3}} }{{2}}, 
\cr 
 t_{\gamma N}^{Sm} \left( {{\bf p}_{\gamma}  ,{\bm \varepsilon}}  \right) = 
 - \frac{{i\ e}}{{4M_{N}} }\left( {{\bm \sigma} \cdot {\bm \varepsilon} \ 
{\bm \sigma} \cdot {\bf p}_{\gamma}  - \left( {1 + \frac{{E_{\gamma} } }{{M_{N}} }} 
\right){\bm \sigma} \cdot {\bf p}_{\gamma}  \ {\bm \sigma} \cdot {\bm \varepsilon}}  \right), 
\cr 
 t_{\gamma N}^{Im} = \mu _{p} \frac{{1 + \tau _{3}} }{{2}} + \mu _{n} 
\frac{{1 - \tau _{3}} }{{2}},\ \ \ 
\cr 
 t_{\Delta N}^{S} \left( {{\bf p}_{\pi} }  \right)\ = \ \frac{{\ 
i\ f_{\Delta N\pi} } }{{M_{\pi} } }\left[ {{\bf p}_{\pi}  \cdot {\bf S}^{ +}  - 
\frac{{1}}{{3}}\left( {1 + C^{\ast} \frac{{E_{\pi} } }{{2M_{N}} }} 
\right)\ {\bm \sigma} \cdot {\bf p}_{\pi}  \ {\bm \sigma} \cdot {\bf S}^{ +} } \right], 
\cr 
t_{\Delta N}^{I} \ = \ {\bm \phi} _{\pi}  \cdot {\bf T}^{ +} .
} 
$$ 
Here, $M_\pi$ is the mass of the pion, ${\bm \sigma}$  is Pauli matrix, 
${\bf S}^{ +}  $ is the operator of the transition between states with 
spin 3/2 and 1/2. 

\section{Results and discussion}
  
Let us consider the properties of the spectator mechanism of isobar 
production using as an example the reaction $^{16}$O$\left( {\gamma ,\ 
\pi ^{ -} pn} \right){}^{14}$O. The procedure for calculating the amplitude 
and cross section of the reaction is in many respects similar to that used 
in Ref. \cite{Glavanakov:2013}. According to the proposed model, in this reaction the 
production of a negative pion is possible as a result of 
two processes with production of the isobar $\Delta ^{0}$ 
\begin{equation}\label{eq4}
\gamma + {}^{16}\mbox{O} \to {}^{14}\mbox{O} + n + \Delta ^{0}
\end{equation}
and the isobar $\Delta ^{ -} $
\begin{equation}\label{eq5}
\gamma + {}^{16}\mbox{O} \to {}^{14}\mbox{O} + p + \Delta ^{ -}.
\end{equation}	
This is followed by the subsequent decays $\Delta ^{0} \to p + \pi ^{ -} $ and $\Delta ^{ - 
} \to n + \pi ^{ -} $. 
\begin{figure}[ph]
\centerline{\includegraphics[width=4.0in]{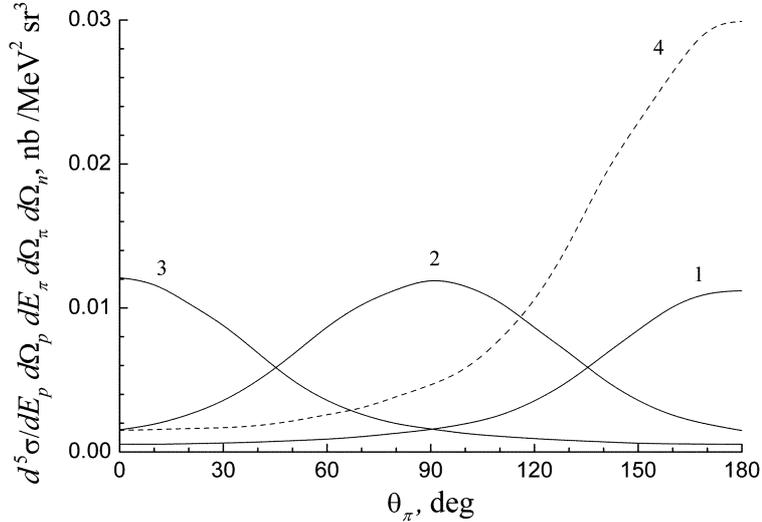}}
\vspace*{8pt}
\caption{\small Differential cross section of the spectator isobar production in the 
$^{16}$O$\left( {\gamma ,\ \pi ^{ -} pn} \right){}^{14}$O reaction 
versus the pion polar angle $\theta _{\pi}  $. Solid curves: $\Delta 
^{0}$-isobar production. Dashed curve: $\Delta ^{ -} $-isobar production. 
Kinematic conditions: $E_{\gamma}  = 500$ MeV, $p_{\pi}  = 170\ $MeV/$c$, 
$p_{p} = 470\ $MeV/$c$, \ $\theta _{n} = 0^{ \circ} ,$ curves 1 and 4: 
$\theta _{p} = 0^{ \circ} ,$ curve 2: $\theta _{p} = 90^{ \circ} ,$ curve 3: 
$\theta _{p} = 180^{ \circ} , \ \phi _{\pi}  = 0^{ \circ} ,$ and $\phi _{p} 
= 180^{ \circ} .$
\protect\label{fig3}}
\end{figure}
Fig. \ref{fig3} shows the differential cross section of the 
${}^{16}$O$\left( {\gamma ,\ \pi ^{ -} pn} \right){}^{14}$O reaction due 
to the spectator mechanism of $\Delta $-isobar production as a function of 
the direction of pion emission. The calculations were performed in the plane 
wave approximation under the following kinematic conditions: $E_{\gamma}  = 
500$ MeV, the neutron emission angle with respect to the photon momentum is 
equal to 0$^{ \circ} $, and the pion and proton momenta are equal to
\begin{equation}\label{eq6}
p_{\pi}  = 170\ \mbox{MeV/}c,\ p_{p} = 470\ \mbox{MeV/}c.					
\end{equation}
The polar angles of proton emission are equal to $0^{ \circ} $ (curves 1 and 
4), $90^{ \circ} $ (curve 2), and $180^{ \circ} $ (curve 3). The solid 
curves are the $\Delta ^{0}$-isobar production cross sections and the dashed 
curve is the $\Delta ^{ -} $- isobar production cross section. As can be 
seen from the figure, the position of the maximum of the 
$\Delta ^{0}$-isobar production cross section is correlated with the 
direction of proton emission. The cross section is maximal when the opening 
angle of the pion and proton is equal to $180^{ \circ} $. On the other hand, 
the position of the maximum of the $\Delta ^{ -} $-isobar production cross 
section is correlated with the neutron emission direction. At the proton 
emission angles equal to $90^{ \circ} $and $180^{ \circ} $, the cross 
section of the reaction $^{16}$O$\left( {\gamma ,\ \pi ^{ -} pn} 
\right)^{14}$O is almost entirely due to the contribution of the $\Delta 
^{0}$-isobar production process (\ref{eq4}). The second process (\ref{eq5}) is suppressed in 
connection with the extremely small Fourier component of the wave functions 
of the bound proton at momentum $\tilde {\bf p}_{n1} = {\bf p}_{n1} - {\bf p}_{\gamma}  $ in 
the kinematic region considered. Thus, the spectator isobar production 
mechanism reproduces the distribution of the cross section by the opening 
angle of the pion and proton at the decay of the $\Delta ^{0}$-isobar weakly 
bound in the $\Delta $-nucleus.

\begin{figure}[t]
\unitlength=1cm
\centering
\begin{picture}(4,4)
\put(-1,0){\includegraphics[width=4cm,keepaspectratio]{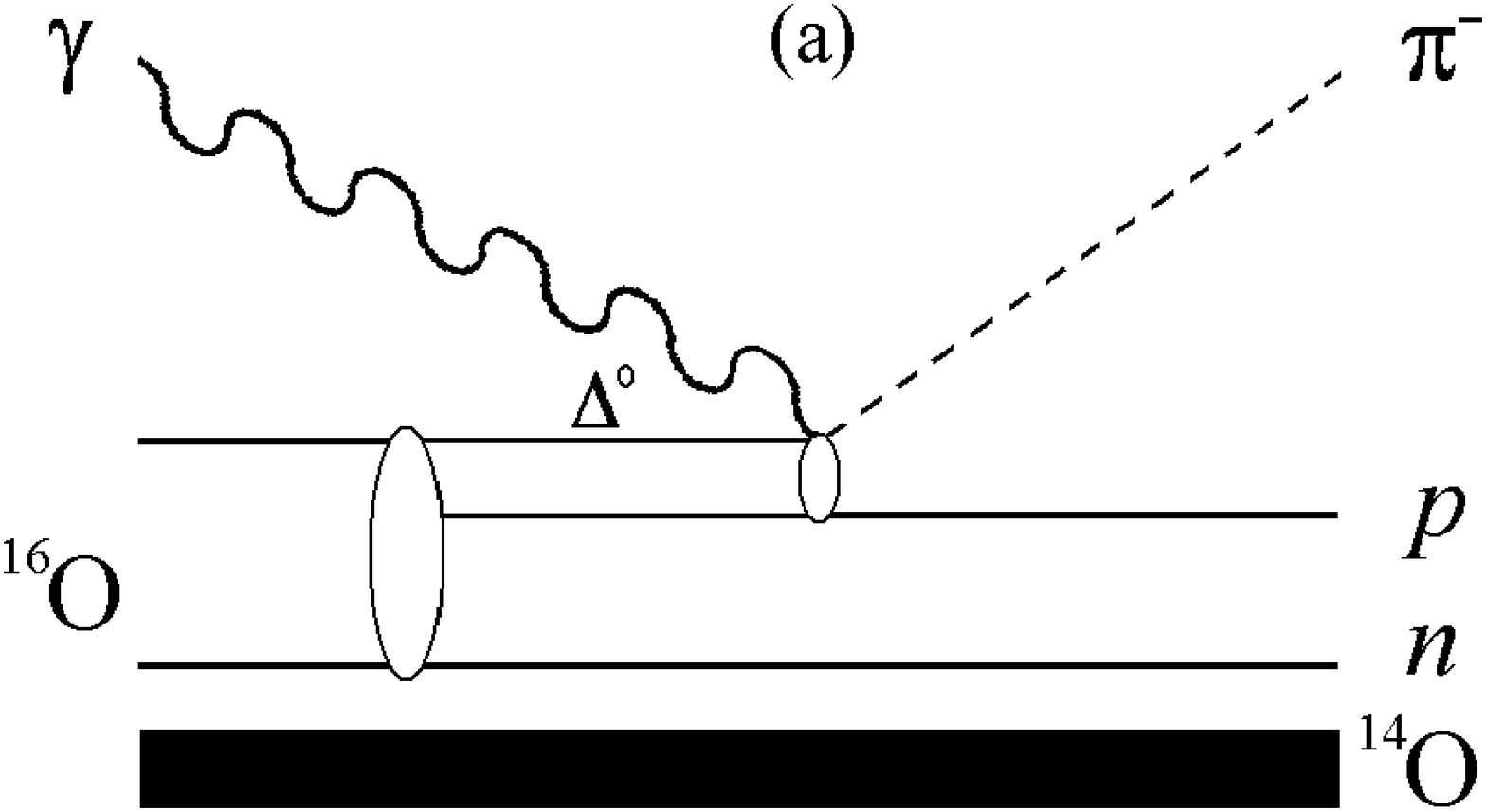}}
\end{picture}
\begin{picture}(4,4)
\put(0,0){\includegraphics[width=4cm,keepaspectratio]{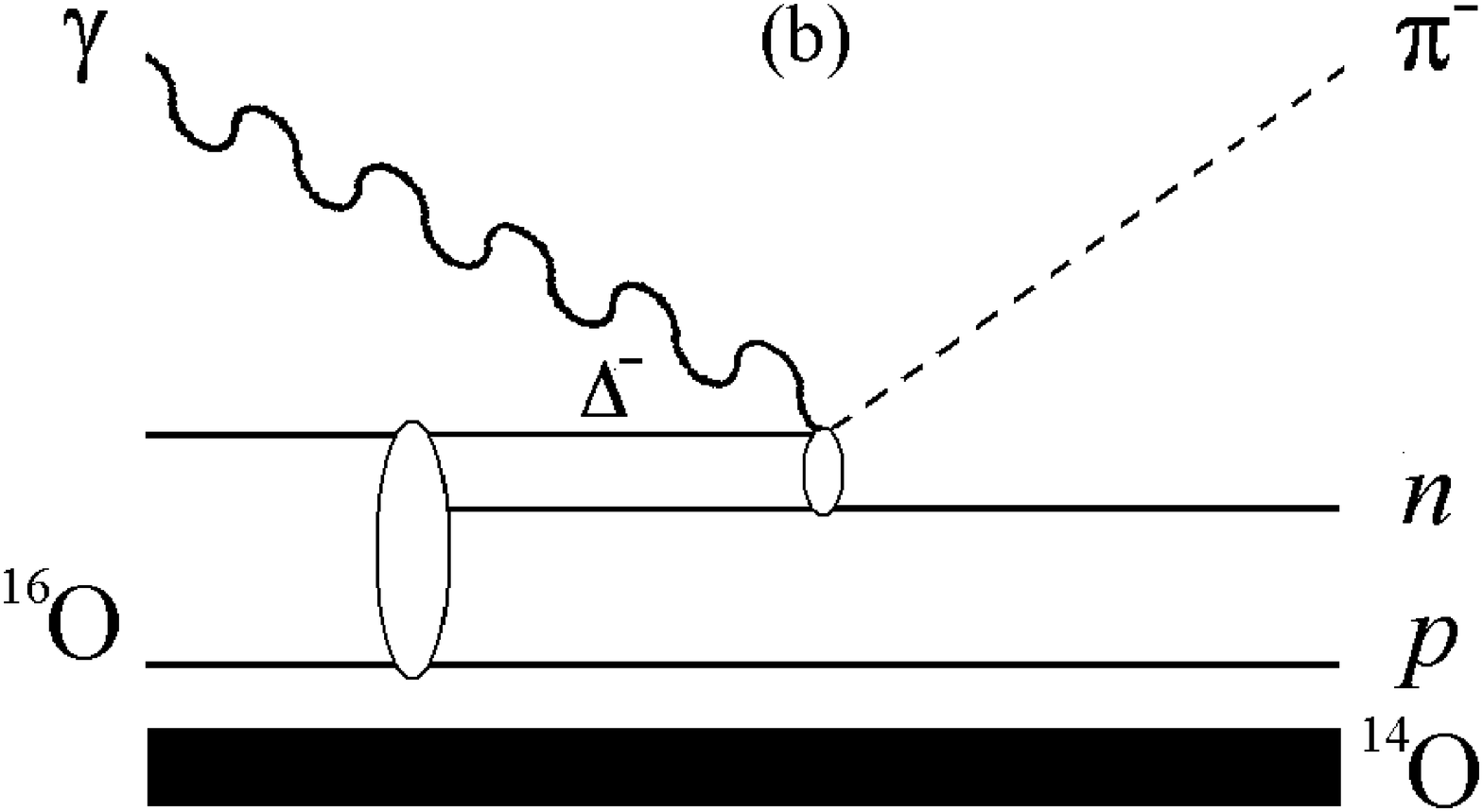}}
\end{picture}
\begin{picture}(4,4)
\put(1,-0.1){\includegraphics[width=4cm,keepaspectratio]{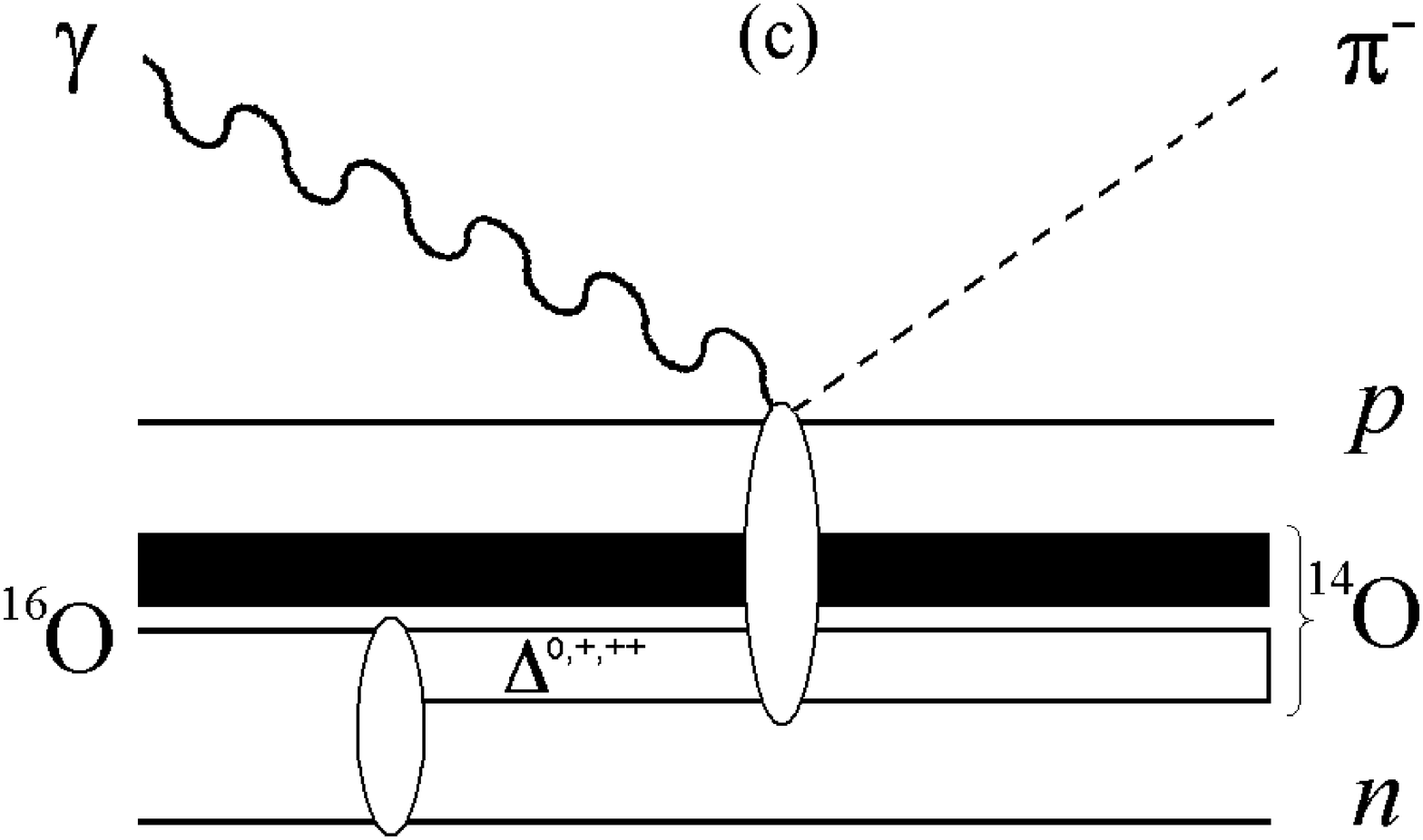}}
\end{picture}
\caption{\small Diagrams illustrating direct mechanisms of the $^{16}$O$\left( {\gamma 
,\ \pi ^{ -} pn} \right)^{14}$O reaction.
\protect\label{fig4}}
\end{figure}

\begin{figure}[h]
\centerline{\includegraphics[width=4.0in]{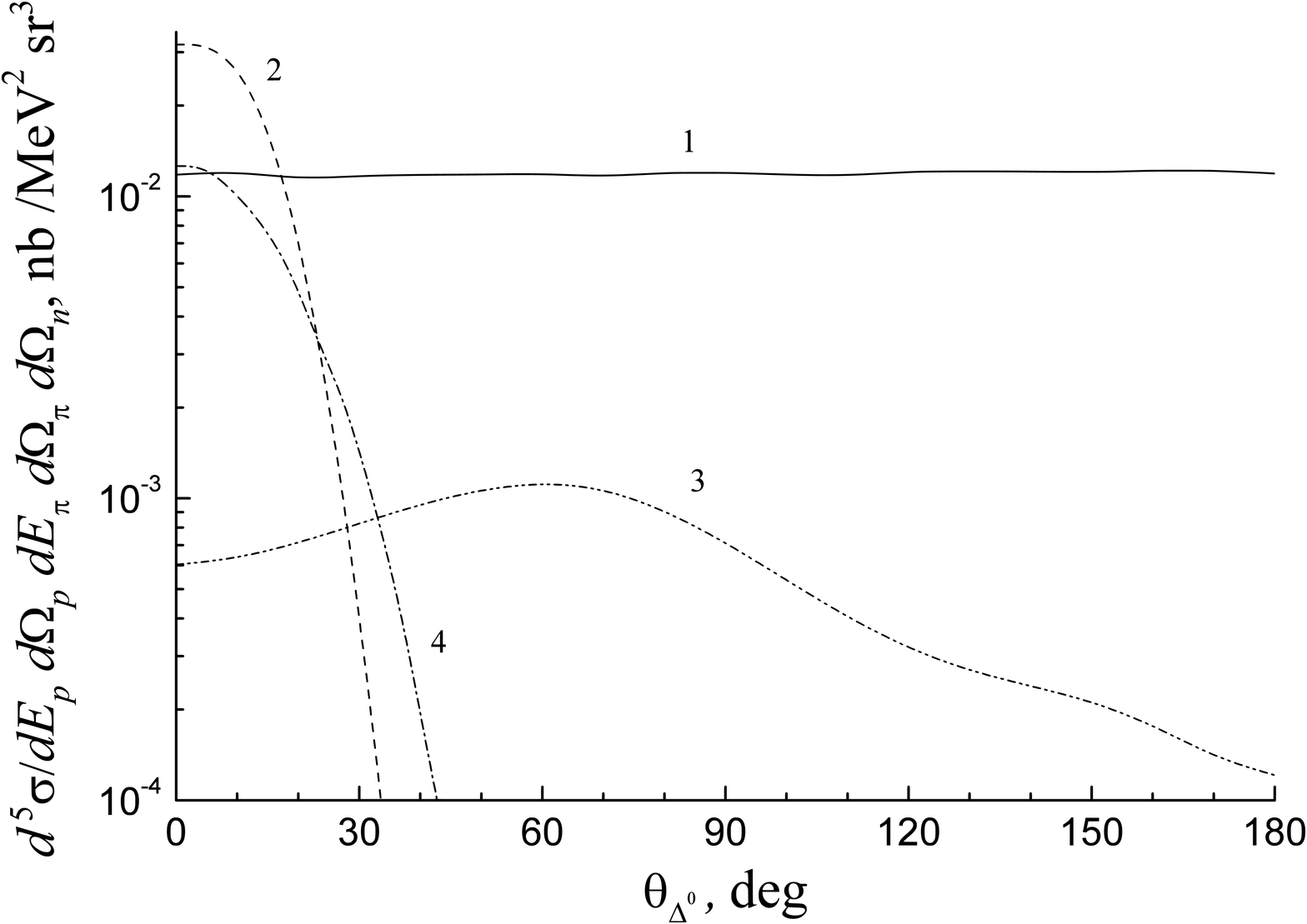}}
\vspace*{8pt}
\caption{\small Differential cross section of the ${}^{16}$O$\left( {\gamma ,\ \pi 
^{ -} pn} \right){}^{14}$O reaction versus the polar angle $\theta _{\Delta 
^{0}} $. Curve 1 and curve 2: contributions of the spectator mechanism of 
$\Delta ^{0}$- and $\Delta ^{ -} $-isobar productions respectively; curve 3: 
contribution of the reaction mechanisms corresponding to the diagrams in 
Figs. \ref{fig4}\textit{a} and \ref{fig4}\textit{b}; curve 4: contribution of the reaction 
mechanism corresponding to the diagram in Figs. \ref{fig4}\textit{c}. Kinematic 
conditions are the same as in Fig. \ref{fig3}, except for $\theta _{\pi}  = 180^{ 
\circ}  - \theta _{p} $ and $\theta _{p} = \theta _{\Delta}  $.
\protect\label{fig5}}
\end{figure}

In the framework of a model that takes into account isobar configurations in 
the ground state of the nucleus \cite{Glavanakov:2013}, pion production is also possible 
through direct $A\left( {\gamma ,\ \pi NN} \right)B$ reaction 
mechanisms. Fig. \ref{fig4} shows diagrams illustrating direct mechanisms of the 
$^{16}$O$\left( {\gamma ,\ \pi ^{ -} pn} \right)^{14}$O reaction, which 
lead to knockout nucleons with momentum sufficiently large for detection in 
experiment using standard methods. In Fig. \ref{fig5} the cross sections of the 
direct and spectator pion productions, depending on the direction of momentum 
${\bf p}_{\Delta ^{0}} = {\bf p}_{\pi}  + {\bf p}_{p} $, are shown. 
Unlike the data in Fig. \ref{fig3}, 
the pion and proton fly in opposite directions: $\theta _{\pi}  = 180^{ 
\circ}  - \theta _{p} $. In this case, for the momenta values (\ref{eq6}), the 
directions of emission of the proton and $\Delta ^{0}$ isobar coincide. Curve 
3 in Fig. \ref{fig5} is the coherent contribution to the cross section of the direct 
reaction mechanisms corresponding to the diagrams in Fig. \ref{fig4}\textit{a} and 
Fig. 4\textit{b}. Curve 4 is the contribution of the diagram in Fig. 
\ref{fig4}\textit{c}. 
The cross sections of the spectator production of the $\Delta 
^{0}$ and $\Delta ^{ -} $ isobars are shown in Fig. \ref{fig5} by curve 1 and curve 2, 
respectively. As can be seen, in the kinematical region considered, the 
$\Delta ^{0}$-isobar production cross section is weakly dependent on $\theta 
_{\Delta ^{0}} $ and dominates at the isobar emission angles, which are 
large $\sim 40^{ \circ} $.

\begin{figure}[ph]
\centerline{\includegraphics[width=4.0in]{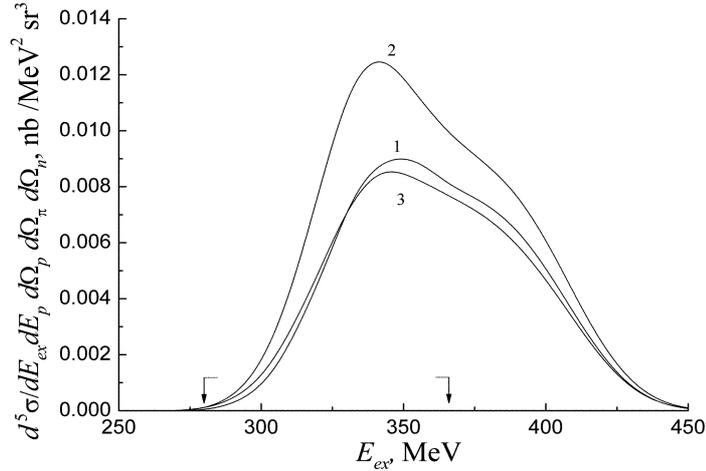}}
\vspace*{8pt}
\caption{\small Differential cross section of the ${}^{16}$O$\left( {\gamma ,\ \pi 
^{ -} pn} \right){}^{14}$O reaction versus energy $E_{ex} $. Kinematic 
conditions: $E_{\gamma}  = 500$ MeV, $p_{p} = 470\ $MeV/$c$,$\theta _{n} = 
0^{ \circ} , \ \theta _{\pi}  = 90^{ \circ} ,$ curves 1: $\theta _{p} = 60^{ 
\circ} ,$ curve 2: $\theta _{p} = 90^{ \circ} ,$ curve 3: $\theta _{p} = 
120^{ \circ} , \ \phi _{\pi}  = 0^{ \circ} , \ \phi _{p} = 180^{ \circ} .$
\protect\label{fig6}}
\end{figure}

An important factor determining the identification of the reaction mechanism 
is the cross section distribution by the invariant mass of the particle 
system in the final state of the nuclear process under study. Fig. \ref{fig6} shows 
the differential cross section  of the reaction ${}^{16}$O$\left( {\gamma 
,\ \pi ^{ -} pn} \right){}^{14}$O as a function of the energy $E_{ex} $, 
which is determined by
\begin{equation}\label{eq7}
E_{ex} = M_{{}^{15}O_{\Delta} }  - M_{{}^{15}O} .						
\end{equation}
Here, $M_{{}^{15}O_{\Delta} }  $ is the invariant mass of the system 
including the residual nucleus $^{14}$O, the proton and negative pion, and 
$M_{{}^{15}O} $ is the mass of the nucleus $^{15}$O. The calculations are 
performed in the region dominated by the spectator mechanism of $\Delta 
^{0}$-isobar production: the neutron emits at an angle $0^{ \circ} $ 
relative to the photon momentum, the polar angle of the pion emission is 
$90^{ \circ} $, the proton emission angles are $60^{ \circ} $ (curve 1), 
$90^{ \circ} $ (curve 2), and $120^{ \circ} $ (curve 3), and the proton 
momentum is 470 MeV/\textit{c}. 
These results can be compared with the empirical data obtained in four 
experiments. In the experiment described in Ref. \cite{Bartsch}, the distribution of the 
cross section by the invariant mass of the pion-nucleon pair and the 
residual nucleus $^{11}$C, produced in the reaction$^{12}$C$\left( {e,e'\pi ^{ 
-} p} \right)^{11}$C, was measured. Narrow peaks in the cross section 
distribution were interpreted as manifestations of the $\Delta $-nucleus 
$^{12}$C$_{\Delta ^{0}} $ in which one neutron is replaced by the $\Delta 
^{0}$-isobar. As a result, the excitation energy of the $\Delta $- nucleus 
$^{12}$C$_{\Delta ^{0}} $, determined similarly to (\ref{eq7}), was estimated.
Based on data from the experiment described in Ref. \cite{Glavanakov:2008}, in which the 
$^{12}$C$\left( {\gamma ,^{}\pi ^{ -} p} \right)$ and $^{12}$C$\left( {\gamma 
,^{}\pi ^{ -} pp} \right)$ reactions were investigated, the excitation 
energy of the $\Delta $-nucleus $^{11}$B$_{\Delta ^{0}} $ was determined. As a 
result of the analysis of the $^{12}$C$\left( {\gamma ,^{}\pi ^{ +} n} 
\right)$ \cite{Liang} and $^{4}$He$\left( {\gamma ,^{}\pi ^{ -} p} \right)$ \cite{Argan} 
reaction data, an estimate for the excitation energy of the $\Delta $-nuclei 
$^{12}$C$_{\Delta ^{ +} } $ and $^{4}$He$_{\Delta ^{0}} $ was obtained \cite{Glavanakov:2009}. In 
Fig. \ref{fig6} an energy interval, in which there are empirical data for the 
excitation energy of $\Delta $-nuclei, is marked by arrows. As 
can be seen, the $E_{ex} $ energy distribution of the cross section of the 
spectator mechanism overlaps with the available data for the $\Delta$-nuclei.

\begin{figure}[t]
\centerline{\includegraphics[width=4.0in]{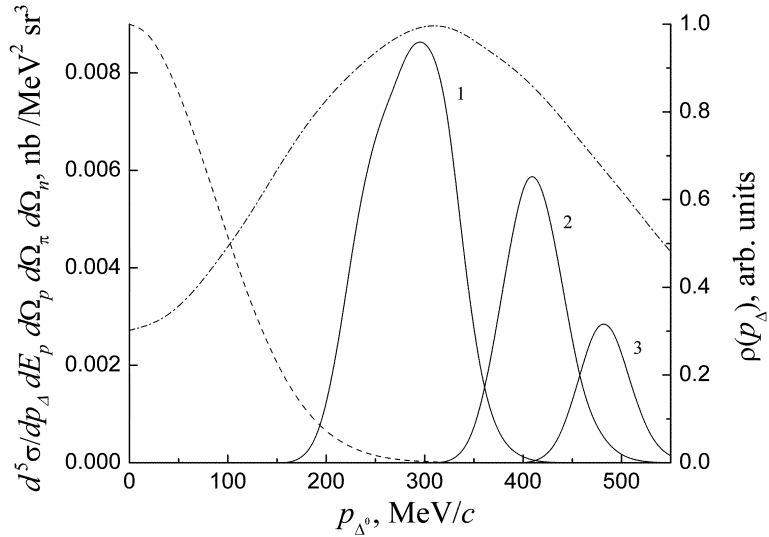}}
\vspace*{8pt}
\caption{\small Momentum distributions (right-hand ordinate) of the virtual isobar in the ground state of 
the nucleus $^{16}$O (dash-dotted curve) and the ``real'' isobar in the $\Delta $-nucleus 
(dashed curve) and differential cross section of the $^{16}$O$\left( {\gamma ,\ \pi 
^{ -} pn} \right)^{14}$O reaction (left-hand ordinate, solid curves) versus the isobar 
momentum $p_{\Delta ^{0}}$. Kinematic conditions 
are the same as in Fig. \ref{fig6}, except for $\theta _{p} = 90^{\circ}$ and 
curve 1: $p_{p} = 470\ $MeV/$c$, curve 2: $p_{p} = 570\ $MeV/$c$, 
curve 3: $p_{p} = 620\ $MeV/$c$.
\protect\label{fig7}}
\end{figure}

In Fig. \ref{fig7} the dependence of the differential cross section of the 
$^{16}$O$\left( {\gamma ,\ \pi ^{ -} pn} \right){}^{14}$O reaction, as a 
function of the isobar momentum $p_{\Delta ^{0}} $, is shown. Within the 
framework of the spectator isobar production model, the dependence of the 
cross section on the momentum $p_{\Delta ^{0}} $ is determined by a number 
of factors, including the momentum distribution of the virtual isobar in the 
ground state of the nucleus and the distribution of the phase space volume 
of the reaction, which limits the range of $p_{\Delta ^{0}} $. From the 
considered properties of the $^{16}$O$\left( {\gamma ,\ \pi ^{ -} pn} 
\right){}^{14}$O reaction, the dependence of the cross section on $p_{\Delta 
^{0}} $ appears to be the most critical feature, which makes it possible to 
identify the spectator mechanism of the reaction and the process of the 
$\Delta $-nucleus production. In Fig. \ref{fig7}, together with the cross sections, 
the momentum distribution of the 
virtual isobar in the ground state of the nucleus $^{16}$O and the momentum 
distribution of the ``real'' isobar in the $\Delta $-nucleus are given. 
The ``real'' isobar momentum 
distribution was obtained under the assumption that the isobar was bound in 
the harmonic oscillator potential in the \textit{s}-state with an oscillator 
parameter $\alpha _{\Delta}  = \alpha _{N} \sqrt {M_{\Delta}  /M_{N}}  $ 
\cite{Klingenbeck}. Here, $\alpha _{N} $ is the nucleon oscillator parameter. The virtual 
transition $NN \to \Delta N$ occurs in the nucleus with greater probability 
at large relative momentum of nucleons, while the capture by the nucleus of 
a real isobar in the bound state is possible at small isobar momentum. These 
circumstances will lead to essentially different dependences of the cross 
sections of these two processes on the total momentum of the pion-nucleon 
pair. It should be noted that the features of the cross section of the 
$A\left( {\gamma ,\ \pi N} \right)$ reaction, interpreted in Ref. \cite{Glavanakov:2009} as 
manifestations of $\Delta $-nuclei, were observed in the kinematic region, 
where the momentum of the pion-nucleon pair takes its minimum value.

\section{Conclusions}

The study of exotic states of nuclei, in which other strongly interacting 
particles besides nucleons are contained in the bound state, is an important 
source of information on hadron interactions. The hypernuclei (nuclei which 
include hyperons) are the most studied at present \cite{Alberico}. The study of 
$\eta$-meson nuclei continues \cite{Machner}, and the $\Delta $-nucleus study is in its 
infancy \cite{Afanasiev}. Most of the information that is supposedly related to $\Delta 
$-nuclei is obtained in experiments, which were performed within the 
framework of another paradigm. Therefore, the results of these experiments 
give rise to more questions than clear answers about the possibility of the 
existence of exotic nucleus states. Based on the opinion that it is first of 
all necessary to make full use of the possibilities of previously tested 
methods for describing nuclear reactions when interpreting experimental 
data, we considered the reaction model caused by isobar configurations in 
the ground state of the nucleus. From the general concepts, the angular 
distributions of the pion-nucleon pairs produced at the decays of the 
isobar-spectator and the isobar, bound in the $\Delta $-nucleus, can be 
similar. Therefore, the aim of this work was to develop a model of the 
spectator isobar production and to study its properties.

An analysis of the spectator mechanism of isobar production is performed 
within the framework of the $\Delta N$-correlation model, which considers 
the isobar and nucleon of the $\Delta N$-system, produced in the nucleus at 
the virtual $NN \to \Delta N$ transition, to be in a dynamic relationship. 
The two-particle transition operator for nuclei $t_{\gamma \pi}  $, was 
obtained by the \textit{S}-matrix approach. A matrix element of the 
transition operator in momentum space followed from the Feynman amplitudes. 
The deduction of the transition operator in configuration space was made 
with the help of the Fourier transformation of each baryon momentum.

The analysis of the spectator mechanism of isobar production is made in the 
kinematic region, where the contribution to the cross section of this 
reaction mechanism is maximal. Numerical estimates of the reaction cross 
section make it possible to draw the following conclusions:

1. The spectator mechanism of isobar production in the $^{16}$O$\left( 
{\gamma ,\ \pi ^{ -} pn} \right){}^{14}$O reaction dominates when the 
isobar emerges at an angle greater than $40^{ \circ} $ with respect to the 
photon momentum;

2. The cross section of the spectator isobar production as a function of the 
opening angle of the pion and nucleon, produced at the isobar decay, has a 
characteristic maximum at an angle of $180^{ \circ} $. This can imitate the 
decay of the bound $\Delta $-isobar;

3. The distribution of the cross section of the spectator isobar production 
over the excitation energy $E_{ex} $ overlaps with the available data for 
the $\Delta $-nuclei;

4. The dependence of the cross section of the $^{16}$O$\left( {\gamma ,\ 
\pi ^{ -} pn} \right){}^{14}$O reaction on the momentum of the pion-proton 
pair is concentrated in the momentum interval which significantly exceeds 
the characteristic momentum of nucleons bound in the nucleus.

\section{Acknowledgements}

The research is funded from Tomsk Polytechnic University Competitiveness 
Enhancement Program grant, Project Number TPU CEP\_PTI\_72$\backslash$2017.

\vspace{4mm}

\nocite{*}  

\bibliography{Paper_Ar}

\end{document}